\documentclass[a4paper,11pt,notitlepage]{article}
\usepackage{a4wide,epsfig,graphics,amsthm,amsfonts,amssymb,amsopn,amsmath,longtable,dcolumn,calc,verbatim,threeparttable,eurosym,lscape,rotating,multirow,threeparttablex, color, setspace, cases}
\usepackage[paper=a4paper,left=30mm,right=30mm,top=33mm,bottom=33mm]{geometry}
\usepackage{natbib, ulem}
\usepackage[latin1]{inputenc}
\usepackage{bbm}
\newcolumntype{.}{D{.}{.}{-1}}
\newcolumntype{d}[1]{D{.}{.}{#1}}

\renewcommand{\baselinestretch}{1.5} 
\theoremstyle{definition}

\theoremstyle{plain}

\newcommand{\indep}{\mbox{$\perp\mspace{-9.0mu}\perp$}}

\newcommand{\bi}{\begin{itemize}}
\newcommand{\ei}{\end{itemize}}

\newcommand{\bc}{\begin{center}}
\newcommand{\ec}{\end{center}}
\newcommand{\bs}{\begin{scriptsize}}
\newcommand{\es}{\end{scriptsize}}
\newcommand{\beq}{\begin{equation}}
\newcommand{\eeq}{\end{equation}}
\newcommand{\ben}{\begin{enumerate}}
\newcommand{\een}{\end{enumerate}}

\DeclareMathSymbol{\R}{\mathalpha}{AMSb}{"52}
\DeclareMathSymbol{\E}{\mathalpha}{AMSb}{"45}

\newcommand{\fe}{\renewcommand{\baselinestretch}{1.32}}


\begin{document}
\renewcommand{\thefootnote}{\fnsymbol{footnote}}

\renewcommand{\baselinestretch}{1.0}

\title{\textbf{Entropy Balancing for Continuous Treatments\footnote{
The author would like to thank Marco Caliendo, Guido Imbens, Martin Lange, Cosima Obst and Sylvi Rzepka as well as participants of the 2019 annual conference of the European Economic Association for helpful comments.
}
} \vspace{0.1cm}}
\author{\textbf{Stefan Tübbicke}\thanks{%
University of Potsdam, e-mail:
\texttt{stefan.tuebbicke@uni-potsdam.de}. \hspace{12.1em} \textcolor{white}{ -} Corresponding address: University
of Potsdam, Chair of Empirical Economics, August-Bebel-Str.\ 89,
14482 Potsdam, Germany. Tel: +49 331 977 3781. Fax: +49 331 977 3210.} \\
} 
\date{\textit{Discussion Paper}\\ This version: \today\\\vspace{-0.5cm}}

\maketitle

\renewcommand{\thefootnote}{\arabic{footnote}} 

\begin{abstract}
	
	\noindent 
	Interest in evaluating the effects of continuous treatments has been on the rise recently. To facilitate the estimation of causal effects in this setting, the present paper introduces entropy balancing for continuous treatments (EBCT) by extending the original entropy balancing methodology of Hainmüller (2012). In order to estimate balancing weights, the proposed approach solves a globally convex constrained optimization problem, allowing for computationally efficient implementation. EBCT weights reliably eradicate Pearson correlations between covariates and the continuous treatment variable. This is the case even when other methods based on the generalized propensity score tend to yield insufficient balance due to strong selection into different treatment intensities. Moreover, the optimization procedure is more successful in avoiding extreme weights attached to a single unit. Extensive Monte-Carlo simulations show that treatment effect estimates using EBCT display similar or lower bias and uniformly lower root mean squared error. These properties make EBCT an attractive method for the evaluation of continuous treatments. Software implementation is available for Stata and R.

	\vspace{5mm}

\end{abstract}
\fe


\thispagestyle{empty}

\newpage
\section{Introduction}
Methods for balancing covariate distributions have become an essential in the tool-kit to control for confounding due to observed covariates. While binary treatments are the most common case encountered in practice, situations in which all units receive some treatment with different intensity or dose are also pervasive in economics and other disciplines. The evaluation of such continuous treatments has gained more attention recently. Examples include the evaluation of job training programs with varying duration \citep{Choe2015, Flores2012, Kluve2012} and subsidies of different magnitude to firms or entire regions \citep{ Egger2012, Bia2007, Mitze2015}. Similar to the binary case, many covariate balancing methods based on the generalized propensity score \citep[GPS,][]{Imbens2000} require an iterative estimation procedure until satisfactory balance is achieved. This is due to the fact that the GPS balances covariates only asymptotically \citep[see][]{Hirano2004, Imai2004, Robins2000}. Therefore, recent developments such as the covariate balancing generalized propensity score \citep[CBGPS,][]{fong2018} and the generalized boosted modeling approach \citep[GBM,][]{Zhu2015} aim to simplify the estimation process by means of algorithmic optimization. 

This paper makes three main contributions to this literature. First, it extends the non-parametric entropy balancing approach by \cite{HM12} for the estimation of balancing weights from the binary treatment framework to the context of continuous treatments. Similar to the original approach, the entropy balancing for continuous treatments (EBCT) algorithm solves a globally convex optimization problem. Balancing weights are obtained by minimizing the deviation from (uniform) base weights subject to zero correlation and normalization constraints. The convex nature of the optimization approach allows for efficient software implementation, converging much faster than other non-parametric balancing methods considered. To facilitate application of the method, software implementation for Stata is provided by the author in the \textit{EBCT} ado-package.\footnote{To install the package, type ``ssc install EBCT, replace'' in the command window.} Implementation for $R$ is also available through the WeightIt package \citep{Greifer2020}. Second, the paper shows that the proposed EBCT method delivers superior finite sample balance in terms of correlations between the treatment variable and covariates in comparison to other re-weighting approaches based on the GPS. In fact, EBCT consistently delivers perfect balance even when other methods tend to fail in this regard due to relatively strong selection into treatment. Moreover, EBCT is more successful in avoiding extreme weights assigned to single units. This property is likely to improve the performance of subsequent effect estimations \citep{Robins2000b, kang2007}. 
Third, extensive Monte-Carlo simulations show that treatment effect estimates based on EBCT do indeed display favorable properties. EBCT yields similar or lower bias and uniformly lower root mean squared error relative to the other methods. 

The remainder of the paper is organized as follows. Section 2 introduces the reader to causal effects of continuous treatments in the potential outcomes framework and necessary identifying assumptions for the consistent estimation in observational studies. Moreover, the section gives a brief overview of the main previous re-weighting methods based on the GPS used as comparisons. Section 3 provides the details of the proposed EBCT method. Section 4 performs several Monte-Carlo simulations  to obtain evidence on the finite sample performance of estimation procedures based on the different balancing approaches. Section 5 then applies the EBCT method to real-world applications and Section 6 concludes.

\section{Causal Effects of Continuous Treatments}

To analyze causal effects in the context of continuous treatments, it is useful to discuss matters in terms of the potential outcomes framework, mainly attributed to  \cite{ROY51} and \cite{RU74}. Following the notation of \cite{Imbens2000} and \cite{Hirano2004}, let us assume that we observe an i.i.d. sample of $N$ individuals $i$ with a vector of pre-treatment covariates $X_i \in \mathbb{R}^K$, where $K$ is the number of covariates. Furthermore, we have information on a post-treatment outcome $Y_i$ and some treatment received with a certain intensity measured by $T_i$ with possible values $\mathcal{T}$. The  potential outcomes are given by $Y_i(t)$  -- also often called the unit-dose response -- denoting the outcome that would have been observed had the unit received treatment with intensity $t$. Aggregating these unit-level responses leads to the dose-response function (DRF) $E[Y_i(t)]$. Along with its derivative  $dE[Y_i(t)]/dt$, the DRF represents the key relationship to be estimated in practice.  If treatment intensities were randomly assigned, comparisons of average outcomes between individuals with different treatment intensities would directly give consistent estimates of these quantities. However, as this is mostly not the case even in experimental settings, the following three identifying assumptions need to be invoked in order to obtain consistent estimates in observational studies.\footnote{Moreover, the outcome model also needs to be specified correctly, calling for flexible functional forms when estimating the DRF parametrically or using non-parametric techniques to model this relationship.}

\paragraph{Identifying Assumptions}

First, conditional on observed pre-treatment covariates $X$, potential outcomes must be independent of the treatment intensity received, i.e.

\begin{equation}
Y_i(t) \indep T_i \mid X_i\ \forall\ t \in \mathcal{T}.
\end{equation}

\noindent This assumption is called the conditional independence assumption \citep[CIA,][]{Lechner2001st} also known as the selection-on-observables assumption \citep{HR85} and requires that the researcher observes all covariates $X$ that simultaneously determine the selection into different treatment intensities as well as the outcome of interest. This is potentially a very strong assumption and needs to be discussed on a case-by-case basis for the application at hand. The second assumption requires there to be common support, i.e. the conditional density of treatment needs to be positive over $\mathcal{T}$:

\begin{equation}
f_{T\mid X}(T=t \mid X_i)>0\ \forall\ t \in \mathcal{T}.
\end{equation}

\noindent If the common support assumption is violated, the sample needs to be trimmed and the DRF is estimated on the subset of observations in order to avoid extrapolation \citep{CHIM09,Lechner2019}. Lastly, one needs to assume the so-called stable-unit treatment value assumption \citep[SUTVA, see][]{RU80}, requiring that each individual's outcome only depends on their own level of treatment intensity. Essentially, this rules out general equilibrium and spill-over effects of treatment \citep[see][for examples]{IW_09, Manski13}.  

While not the focus of this paper, the estimation of effects of continuous treatments may be combined with other identification approaches than selection-on-observables. For example, methods described may be applied to the estimation of treatment effects in a conditional Difference-in-Differences setting \citep{AB05} where all units are affected by some natural experiment to a different degree. Alternatively, estimating DRFs of continuous instrumental variables \citep{AIR96} that are only valid conditional on covariates is likely to expand knowledge about which units are actually induced to receive some treatment by the instrument.

\paragraph{(Re-weighting) Methods based on the Generalized Propensity Score}\label{GPS}

Non-parametric estimation of DRFs by comparing outcomes of individuals with exactly the same set of $X$ but different $T$ quickly becomes infeasible with growing dimension of $X$. To avoid this curse of dimensionality, \cite{RR83} show for the binary treatment case that it is sufficient to condition on the scalar propensity score instead of the multidimensional vector $X$ in order to control for confounding due to observed covariates. Similarly, \cite{Hirano2004} show that the conditional independence assumption also holds by conditioning on the generalized propensity score (GPS) $R=f_{T\mid X}(T \mid X)$, i.e. the conditional density of the treatment intensity evaluated at $T$ and $X$. In order to estimate the GPS, \cite{Hirano2004} assume the treatment follows a normal distribution and perform an Ordinary Least Squares (OLS) regression of $T$ on $X$ and obtain the GPS as

\begin{equation}
\hat{R}_i=\frac{1}{\sqrt{2\pi \hat{\sigma}}} exp \left\{-\frac{1}{2\hat{\sigma}^2} (T_i - \hat{\beta}' X_i)^2 \right\},
\end{equation}

\noindent where $\hat{\beta}$ is the regression coefficient vector and $\hat{\sigma}$ is the standard error of the disturbance term. Based on this estimated GPS, \cite{Hirano2004} advocate estimating the DRF by controlling for the GPS via flexible parametric regression.\footnote{Software is provided by \cite{BM08} in the \textit{doseresponse} ado-package for STATA.} 

As conditioning on the \textit{correctly specified} GPS balances $X$ across different levels of $T$ only in expectation, (iteratively) checking covariate balance is a necessary step in the estimation of DRFs.  \cite{Hirano2004} suggest to do this by conducting GPS-adjusted $t$-tests on the equality of means across strata in $T$. As this can be somewhat cumbersome and the stratification may lead to information loss \citep{Austin2019}, a different strand of literature followed the idea of estimating balancing weights instead. These weights allow to directly assess the resulting balancing quality by comparing (absolute) Pearson correlations in the raw data and in the re-weighted sample.\footnote{The sub-classification approach by \cite{Imai2004} faces similar issues and is therefore not discussed at this point.} One such approach -- originating from inverse probability weighting \citep[IPW,][]{Horvitz52} -- is provided by \cite{Robins2000} who generalize IPW and show that weights defined as 

\begin{equation}\label{balweight}
w_i=\frac{f_T(T_i)}{f_{T\mid X}(T_i \mid X_i)} 
\end{equation}

\noindent render the treatment intensity uncorrelated with covariates in the re-weighted sample in expectation. Similar to \cite{Hirano2004}, \cite{Robins2000} estimate (un-) conditional densities $f_T(T)$ and $f_{T\mid X}(T \mid X)$ based on OLS regressions and the normality assumption.\footnote{For an assessment of the performance of other IPW methods for continuous exposures with different distribution assumptions and estimation approaches, see \cite{Naimi2014}.} 

With the goal of avoiding iterative balance-checking, re-specification and estimation of the GPS, \cite{fong2018} generalize the covariate balancing propensity score (CBPS) methodology of \cite{Imai2014} to include continuous treatments. For their parametric approach -- henceforth covariate balancing generalized propensity score (CBGPS) -- they derive the parametric structure of balancing weights based on the normality assumption. Parameters are estimated using the generalized method of moments by minimizing squared Pearson correlations between the treatment and covariates in the re-weighted sample.\footnote{\cite{Huffman2018} further generalize the CBGPS approach of \cite{fong2018} to allow for time-varying interventions.} Moreover, \cite{fong2018} also provide a non-parametric version -- denoted as npCBGPS for the remainder of the paper. This approach obviates the need to specify a parametric structure for balancing weights by maximizing the empirical likelihood \citep{owen1990, Qin1994} subject to imbalance constraints. The imbalance constraints are chosen to allow for some finite sample imbalance in order to improve the convergence properties of the proposed algorithm.\footnote{The pre-specified degree of imbalance in these constraints is left as a tuning parameter for the researcher. For the purpose of this paper, the tuning parameter will be left at its pre-specified level.} Compared to the CBGPS, the non-parametric approach is likely to come with a computational cost which may be quite substantial for datasets with a large number of observations and/or covariates that need to be balanced. 

Another non-parametric approach for the estimation of balancing weights is provided by \cite{Zhu2015}. Their approach is based on machine-learning techniques and adapts  generalized boosted models (GBM) of \cite{Mccaffrey2005} to the context of continuous treatments. GBM uses a boosting algorithm to estimate the GPS, plugging resulting estimates into equation (\ref{balweight}). Balance in terms of absolute Pearson correlations is optimized via the number of regression trees grown.\footnote{For simulations, the maximum number of titerations is set to 20,000 with a shrinkage of 0.05\%.}  

As will become clear in the remainder of the paper, existing automated balancing approaches based on the GPS mostly improve upon balancing quality relative to IPW. However, the re-weighting procedures tend not to achieve satisfactory balance when selection into different treatment intensities is relatively strong, leaving estimates susceptible to bias due to residual imbalance. The next section provides a solution to this issue by extending the entropy balancing method for binary treatments by  \cite{HM12} to the context of continuous treatments.

\section{Extending the Entropy Balancing Scheme}
The original entropy balancing (EB) method by \cite{HM12} is a non-parametric pre-processing tool to estimate balancing weights for binary treatments, i.e. it re-weights control units to exactly match pre-specified covariate moments of the treatment group. The convex nature of the optimization problem solved by EB guarantees excellent balancing properties of resulting weights. Moreover,  \cite{Zhao2017} show that  EB is doubly-robust \citep{Robins1995} and that it reaches the semi-parametric efficiency bound derived by \cite{Hahn1998}. These properties make EB an attractive candidate for the extension to the context of continuous treatments in order to improve upon existing balancing approaches. The remainder of this section introduces the proposed entropy balancing for continuous treatments (EBCT) approach. 


\paragraph{Entropy Balancing for Continuous Treatments}

For notational convenience, assume that the treatment intensity and covariates are standardized to mean zero. Furthermore, define the column vector $g(T_i,X_i)=[T_i, X_i^T,T_iX_i^T]^T$. The EBCT method aims to solve the following constrained minimization problem:

\begin{equation} \label{minimization}
 \begin{aligned}
       \min_{w} H(w)= \sum_{i=1}^N h(w_i) \quad s.t.\quad   \sum_{i=1}^N w_i g(T_i,X_i) &=&0\\
\sum_{i=1}^N w_i &=&1\\
w_i &>& 0\  \forall i
	       \end{aligned}
\end{equation}

\noindent EBCT minimizes the loss function $H(w)$ subject to the balancing constraints and the normalizing constraints that weights have to sum up to one and be strictly positive. Weights that satisfy (\ref{minimization}) retain unconditional means of covariates as well as the treatment variable and most importantly, they purge the treatment variable from its correlation with covariates. The inclusion of higher-order or interaction terms in the list of covariates allows the researcher to achieve balance not just regarding the mean of covariates, but also regarding higher and cross-moments. Compared to the original EB method, the optimization problem (\ref{minimization}) differs in terms of balancing constraints imposed and the set of units for which balancing weights are being estimated. Essentially, EBCT re-weights \textit{all units} to achieve \textit{zero correlations} between the treatment variable and covariates.

\paragraph{Implementation} 
In oder to implement the EBCT approach, one needs to decide upon the loss function $H(w)$. Following \cite{HM12}, this paper uses a loss function based on the \cite{Kullback1959} entropy metric $h(w_i)=w_i ln(w_i/q_i)$, where $q_i$ are some base weights chosen by the analyst. If no base weights are specified, uniform weights $q_i=1/N\ \forall\ i$ are used. This implies that EBCT chooses balancing weights such that they differ as little as possible from baseline weights while achieving zero correlation in the re-weighted sample. Notice that the loss function attains a minimum at $w_i=q_i\ \forall\ i$ and is undefined for non-positive weights. The latter property allows to drop the positivity constraint on weights, reducing the optimization problem to one with only equality constrains. Using the Lagrange method, the constrained optimization can be re-written as an unconstrained optimization as

\begin{equation}\label{lagrange}
\min_{w,\lambda,\gamma} \mathcal{L}(w,\lambda,\gamma)=\sum_{i=1}^N w_i ln(w_i/q_i)-\lambda \left\{ \sum_{i=1}^N w_i - 1\right\} - \gamma^T \left\{\sum_{i=1}^N w_i g(T_i,X_i) \right\}, \\
\end{equation}

\noindent where $\lambda$ and $\gamma$ are Lagrange-multipliers on the constraints. As $\partial^2 \mathcal{L}/\partial w_i^2>0$ for all $w_i>0$ and because the constraints are linear in $w_i$, the optimization problem (\ref{lagrange}) has a global minimum if the constraints are consistent \citep[][chapter 5]{boyd2004}. In order to reduce the dimensionality of the optimization problem, the implied structure of balancing weights is obtained by re-arranging the first-order condition $\partial \mathcal{L}/ \partial w_i=0 $ and plugging the result into  the condition $\partial \mathcal{L}/ \partial \lambda=0$. This yields the weighting function in terms of the Lagrange-multipliers $\gamma$, $q_i$ and $g(T_i,X_i)$ as

\begin{equation} \label{weights}
 \begin{aligned}
       	w_i &=&   \frac{q_i\ exp\left\{\gamma^T g(T_i,X_i)\right\}}{ \sum_{i=1}^N q_i\ exp\left\{\gamma^T g(T_i,X_i)\right\}},\\
	       \end{aligned}
\end{equation}
\vspace{0.1em}

\noindent where $\lambda$ has been cancelled out. Hence, weights implied by EBCT are a log-linear function of a linear index containing covariates, the treatment intensity and their cross-products. Substituting this expression into the Lagrange function yields the dual $ \mathcal{L}^d$ as 

\begin{equation}\label{lagrange_dual}
\mathcal{L}^d(\gamma)=- ln\left( \sum_{i=1}^N q_i\ exp\left\{\gamma^T g(T_i,X_i)\right\}  \right). \\
\end{equation}

Differentiating $ \mathcal{L}^d$ with respect to $\gamma$ yields the $2K+1$ first-order conditions in $2K+1$ unknowns

\begin{equation} \label{moments}
\left\{
 \begin{aligned}
	&  \sum_{i=1}^N  \frac{exp\left\{{\gamma^*}^T g(T_i,X_i)\right\} g(T_i,X_i)}{ \sum_{i=1}^N exp\left\{{\gamma^*}^T g(T_i,X_i)\right\}}\\
       \end{aligned}
 \right\} = 0,
\end{equation}
\vspace{0.4em}

\noindent where $ {\gamma^*}$ refer to the multiplier values at the optimum.\footnote{Note that first-order conditions have been multiplied by -1. In comparison, finding the optimum of (\ref{lagrange}) requires choosing $N+2(K+1)$ parameters in total.} As equations (\ref{moments}) are non-linear in those multipliers, they have to be solved for numerically. This is done using a quasi-Newton optimization approach. Due to the convexity of the optimization problem, the algorithm tends to converge much faster than GBM and npCBGPS, especially in large datasets.\footnote{Computation times for the estimation of balancing weights using the approaches described are given in section 4.} Once values for $ {\gamma^*}$ are obtained, balancing weights are backed out using (\ref{weights}) for subsequent analysis. As noted by \cite{HM12}, optimization can be performed iteratively to limit the influence of units with potentially extreme weights. To do so, the researcher estimates EBCT weights and truncates excessive weights beyond some threshold, e.g. 4\% as suggested by \cite{IB04}. Then, the estimation is repeated with truncated weights as base weights. Resulting weights still lead to finite sample balance but display smaller maximum weights.
 
\section{Monte-Carlo Simulations}\label{mc}
In this section, the finite sample properties of weighting approaches in terms of balancing outcomes as well as resulting effect estimates are compared using Monte-Carlo simulations.\footnote{Because the \cite{Hirano2004} and the \cite{Imai2004} procedures do not allow to directly assess covariate balance in terms of correlations, their approaches are excluded from the comparison. See also \cite{Austin2018} for additional evidence on finite sample performance of existing estimators based on the GPS.}  In general, the simulation design is chosen to mimic relevant features of datasets encountered in empirical practice and is similar in spirit to the design by \cite{HM12}. As such, the design is built around a variety of different covariate distributions. In general, the treatment intensity is modeled as a linear function of covariates and the outcome is specified as (non-) linear and (non-) additive function of the covariates with a constant additive treatment effect. In total, 18 different scenarios are constructed by altering the degree of non-linearity and non-additivity in the outcome equation, the degree of selectivity in the selection equation as well as the degree of mis-specification of balancing weights. Moreover, sample sizes are varied with $N=200,\ 500$ and $1,000$. For each simulation scenario and sample size, $R=1,000$ independent replications are performed. As simulation results are quite similar across sample sizes, only results for $N=200$ are presented in the main text. Additional results for larger sample sizes can be found in Appendix A.

\paragraph{Simulation Design}
Each simulated dataset consists of ten (partially correlated) covariates $X_1,...,X_{10}$ entering the selection equation: $X_1 \sim U[0,5]$, $X_2 \sim \chi^2_2$, $X_3$ to $X_5$ are binary indicators based on one underlying standard normal variable with cut-offs of ($-\infty$,-1], (-1, 0] and (0,1]\footnote{The interval (1,$+\infty$) serves as a reference category with a coefficient of zero.}, $X_6\sim \mathcal{B}(p=0.5)$ and $X_7$ to $X_{10}$ are jointly standard normal with covariance of 0.2. The treatment equation is specified as

\begin{equation}\label{select}
\begin{split}
T_i=X_{1i}+0.6 X_{2i}+1.2 X_{3i} + X_{4i} + 0.5 X_{5i} + X_{6i}\\
 + 0.8 X_{7i} + 0.8 X_{8i} + 0.8 X_{9i} +0.8 X_{10i} + \sigma \varepsilon_i , 
\end{split}
\end{equation}

\noindent where $\varepsilon$ is a standard normal error term and $\sigma$ is the scale parameter governing the standard deviation of the composite error term.  Based on this equation, moderate selectivity into treatment is generated with $\sigma=4$ and strong selection is obtained by setting $\sigma=2$.\footnote{While these labels are obviously quite arbitrary, values of $\sigma$ have been chosen to roughly mirror the selectivity patterns of the empirical applications presented in section 4.} To investigate the robustness of the estimation approaches to mis-specification of the selection equation, three different specifications for the estimation of balancing weights are assumed:

\bi
\item Specification 1:  \quad  $\hat{E}[T\mid X]=\alpha_0+ \sum_{k} \alpha_k X_{k}$
\item Specification 2: \quad  $\hat{E}[T\mid X]=\alpha_0+ \sum_{k\notin \{1,7\}} \alpha_k X_{k} + \alpha_1 \sqrt{X_1} + \alpha_7 X_7^2$
\item Specification 3: \quad $\hat{E}[T\mid X]=\alpha_0+ \sum_{k\notin \{1,2,7\}} \alpha_k X_{k} + \alpha_1 \sqrt{X_1} +  \alpha_2 X_8^2 +\alpha_7 X_7^2$.
\ei

\noindent Specification one is correctly specified and is thus expected to yield the lowest bias and root mean squared error (RMSE). Specification two introduces mild mis-specification via mis-measurement of $X_1$ and $X_7$. Finally, specification three further increases bias by falsely omitting $X_2$ from the specification, replacing it by $X_8^2$ instead.

\vspace{1em} \noindent The outcome equation is modeled as

\begin{equation}\label{outcome}
Y_i= (X_{1i}+X_{2i})^\eta+   X_{5i} +  X_{6i} +  X_{7i} + T_i + \xi_i , 
\end{equation}  

\noindent where $\xi \sim N(0,5)$. The degree of non-linearity and non-additivity is chosen via $\eta$. Three scenarios are considered. In outcome design one, $\eta$ is set to one, yielding a linear specification of $Y$ in $X$. For mild deviations from linearity and additivity in covariates in design two, $\eta=1.25$ is used and moderate non-linearity/additivity is obtained by setting $\eta=1.5$ for design three. In all designs, a linear DRF with a derivative of one is assumed. In each simulation replication, balancing weights are estimated in a first step and second, a weighted simple linear regressions of the outcome on the treatment intensity is performed in order to obtain estimates of the treatment effect. 

If the selection equation is correctly specified and balancing weights achieve zero correlations between the treatment variable and covariates, estimates are expected to be unbiased. Under mis-specification, all estimators ought to yield biased estimates which is likely to be exacerbated by the degree of non-linearity/additivity in the outcome equation.

\begin{figure}[ht!] 
\caption{Simulation Results - Balancing Properties} 
\label{bal_sim}
\bc
\includegraphics[scale=0.6, clip, trim=0cm 0cm 0cm 0.5cm]{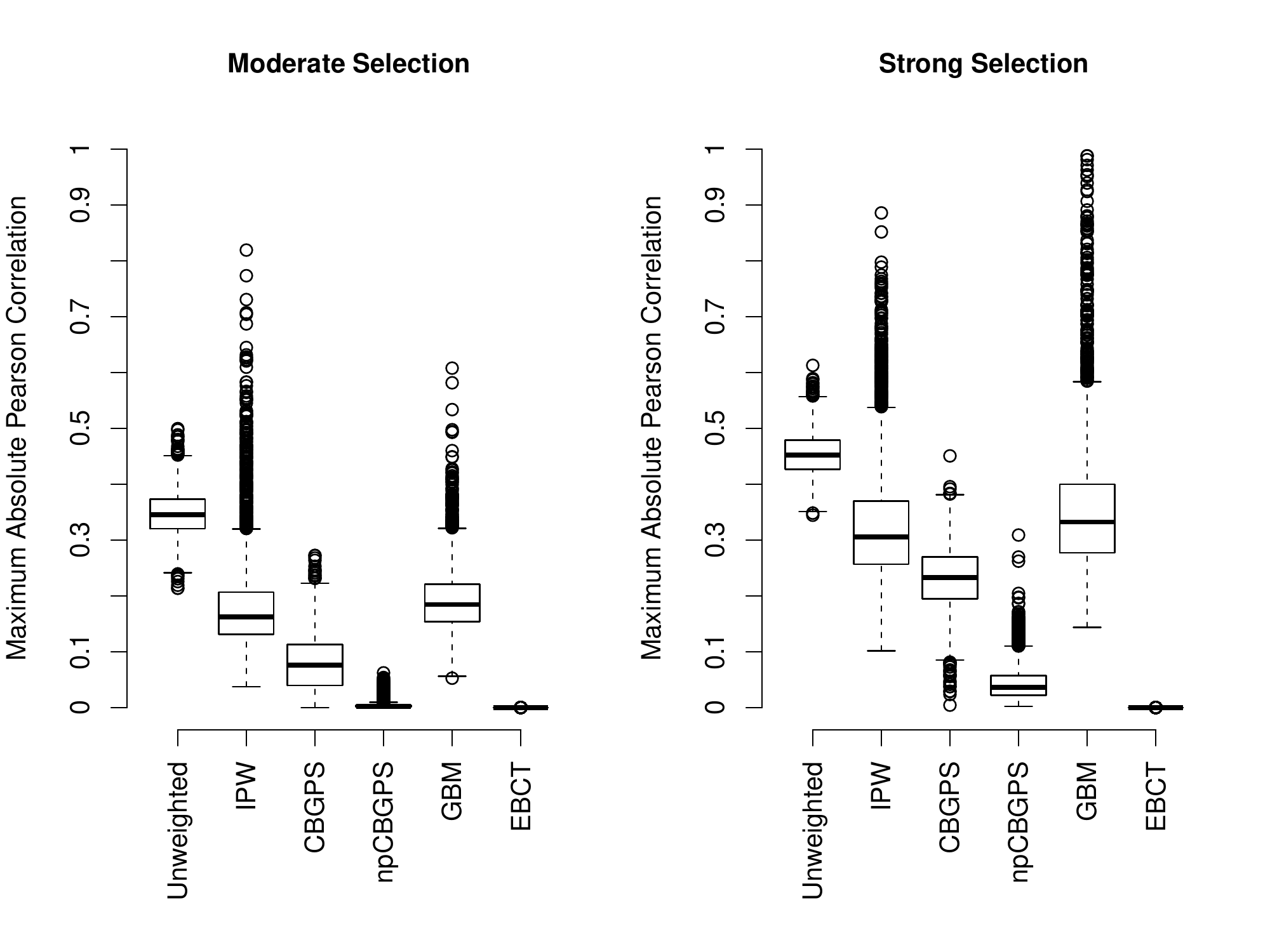}
\begin{scriptsize}
\begin{spacing}{1.0}
\parbox{42em}{Note: This graph plots the distribution of the maximum absolute correlation coefficients between the treatment intensity $T$ and the covariates $X$ for Monte-Carlo simulation designs with correctly specified $E[T\mid X]$, split by the degree of selection into treatment, both in the raw sample (\textit{unweighted}) as well as in re-weighed samples. Re-weighting approaches employed are  inverse probability weighting estimated via OLS \citep[IPW, see][]{Robins2000}, (non-) parametric covariate balancing generalized propensity scores \citep[np-/CBGPS, see][]{fong2018}, generalized boosted modeling \citep[GBM, see][]{Zhu2015}  as well as the novel entropy balancing for continuous treatments (EBCT).} \vspace{-9em}
\end{spacing}
\end{scriptsize}
\ec
\end{figure}

\paragraph{Balancing Quality} 
First and foremost, the different estimation procedures aim to balance covariates across different treatment intensities. Hence, an important criterion regarding the empirical performance of these procedures is the degree to which they actually deliver finite sample balance.  Simulation results on the distribution of balancing quality indicators for both the moderate and strong selection into treatment under correct specification can be found in the left and the right panel of Figure \ref{bal_sim}, respectively. Results for the mis-specified cases are not presented as they yield the same conclusions. In the spirit of \cite{DS13}, the largest absolute Pearson correlation coefficient is used as a balancing indicator, putting more focus on the least balanced covariate instead of the average balancing quality, recognizing that even small imbalances may lead to substantial bias if the covariate is a strong predictor of the outcome.

\bc
[Insert Figure \ref{bal_sim} about here]
\ec

\noindent When selection into treatment is moderate, i.e. initial maximum absolute Pearson correlations are around 35\%, all methods tend to improve balancing in the re-weighted dataset to some degree. IPW tends to result in highly variable balancing quality and sometimes even leads to an increase in imbalance. Similar results are obtained for the GBM approach. The parametric CBGPS reduces maximum correlations much closer towards zero but still displays substantial variability in balancing outcomes. The non-parametric CBGPS approach outperforms its parametric counterpart by consistently delivering correlations near zero in the re-weighted data when selection into treatment is moderate.  However, when selection into treatment is strong, i.e. when initial absolute correlation attain a maximum of around 45\%, the simulation results show that now even npCBGPS yields more variable balancing quality, frequently surpassing the 0.1 rule-of-thumb threshold proposed by \cite{Zhu2015}. While balancing quality tends to deteriorate with initial imbalance for the other approaches, EBCT effectively eliminates correlations in the re-weighted simulation data independent of the magnitude of initial correlations.

\begin{figure}[ht!] 
\caption{Simulation Results - Weight Distributions} 
\label{weight_sim}
\bc
\includegraphics[scale=0.6, clip, trim=0cm 0cm 0cm 0.5cm]{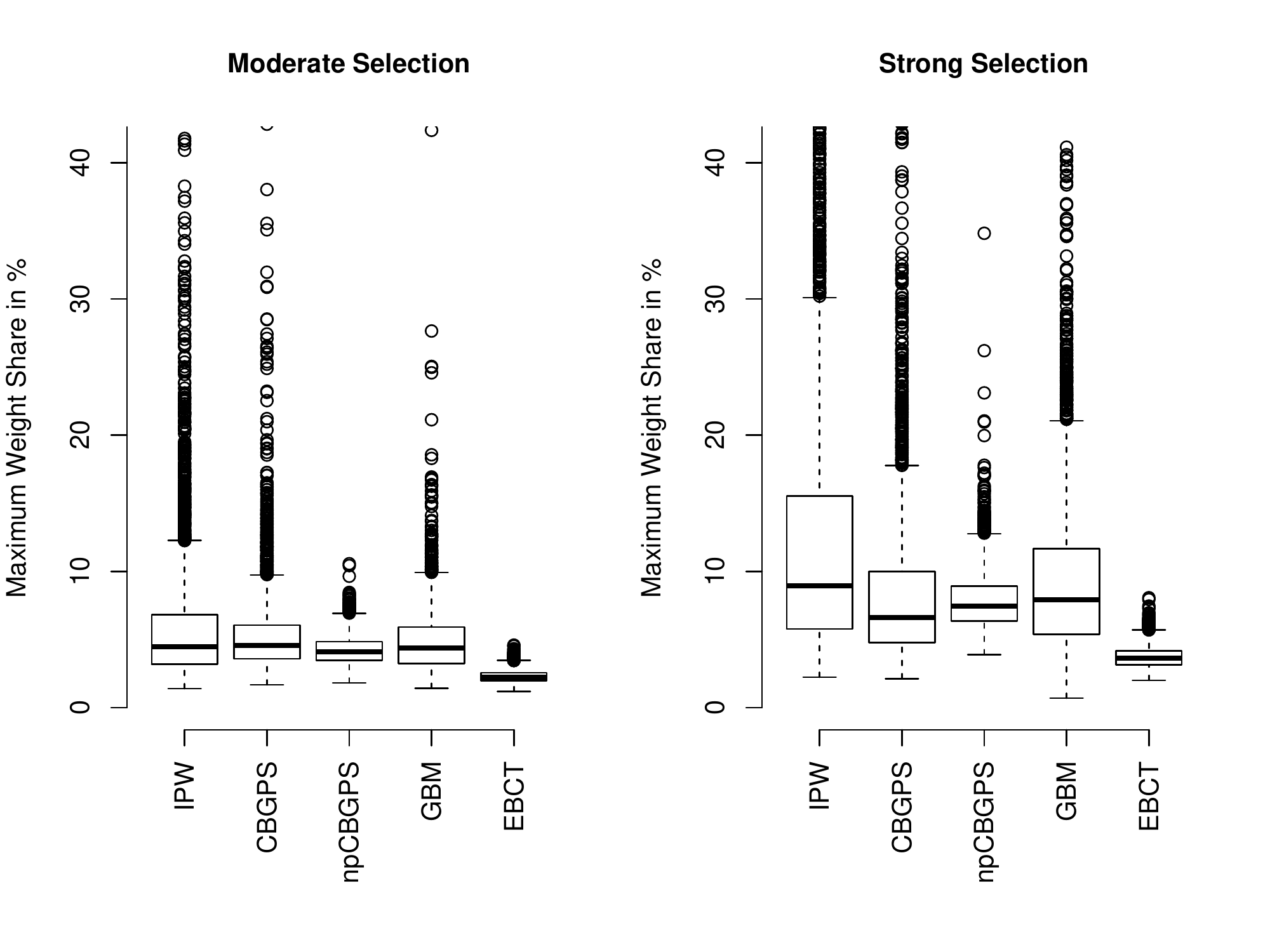}
\begin{scriptsize}
\begin{spacing}{1.0}
\parbox{42em}{Note: This graph plots the distribution of the maximum weight shares for Monte-Carlo simulation designs with correctly specified $E[T\mid X]$, split by the degree of selection into treatment in re-weighed samples. Re-weighting approaches employed are  inverse probability weighting estimated via OLS \citep[IPW, see][]{Robins2000}, (non-) parametric covariate balancing generalized propensity scores \citep[np-/CBGPS, see][]{fong2018}, generalized boosted modeling \citep[GBM, see][]{Zhu2015}  as well as the novel entropy balancing for continuous treatments (EBCT). } \vspace{-0.5em}
\end{spacing}
\end{scriptsize}
\ec
\end{figure}

\paragraph{Distribution of Balancing Weights} 
While the balancing quality of weights is certainly an important criterion, it may be the case that finite sample balance comes at the cost of overly large weights for just a few units when estimating treatment effects. This is likely to substantially reduce the performance of resulting estimates and should be avoided \citep{Robins2000b, kang2007}. To provide some evidence of the performance of the balancing approaches in this regard, Figure \ref{weight_sim} displays the distribution of the maximum weight share held by a single unit across all simulations with correctly specified balancing weights, again split by the degree of selection into treatment.\footnote{The graph only shows maximum weights up to a value of 40\% in order to enhance visibility of the weight distribution. Especially IPW and GBM often lead to larger weight shares in the case of strong selection into treatment. Similar to the analysis of balancing quality, an analysis based on the mis-specified $E[T\mid X]$ gives rise to the same conclusions.}

\bc
[Insert Figure \ref{weight_sim} about here]
\ec

The results suggest that maximum weight shares for IPW, the parametric and the non-parametric CBGPS as well as GBM are fairly similar most of the time, although IPW, the parametric CBGPS and GBM display more variability and produce extreme weights more frequently. EBCT delivers smaller maximum weight shares on average and displays much less variability in maximum weights compared to the other approaches. Similar to the balancing criterion, the relative advantage of EBCT grows with the degree of selection into treatment. 

\paragraph{Bias and Root Mean Squared Error} 

Turning to the finite sample properties of effect estimates based on the re-weighting approaches, Table \ref{mc_res} compares Monte-Carlo results on absolute bias and the root mean squared error (RMSE).

\bc
[Insert Table \ref{mc_res} about here]
\ec

As expected, mis-specification of selection equation generally increases bias and RMSE for all estimators. Similarly, increases in the degree of non-linearity/additivity in the outcome equation tend to lead to larger bias and RMSE. Regarding the individual performance of estimators, IPW, the parametric CBGPS and GBM tend to display the highest absolute bias and RMSE. The non-parametric CBGPS reduces bias and RMSE compared to its parametric counterpart in all simulation scenarios.  However, when faced with strong selection into treatment the npCBGPS yields biased estimates even when the selection equation is correctly specified. If the outcome is sufficiently non-linear and non-additive in covariates, bias is substantial. EBCT on the other hand consistently delivers essentially unbiased estimates even when selection into treatment is strong as long as all relevant and correctly measured covariates are included in the specification of balancing weights. Moreover, EBCT yields the lowest RMSE across all simulation scenarios independent of whether the selection equation is correctly specified or not. 

\section{Empirical Applications}

In this section, the EBCT methodology and comparison methods are applied to the estimation of dose-response functions using real data from two well-known examples on the size of lottery winnings and labor earnings as well as smoking intensity and medical expenditures. An additional empirical example on the evaluation of a place-based development subsidy is presented in Appendix B. Note that the analysis in this section remains agnostic about the validity of the conditional independence assumption and hence, estimates are only interpreted as conditional associations. For simplicity, outcome regressions are performed via weighted least squares regression based on the estimated balancing weights using a cubic polynomial in the respective treatment variable. From these estimates, the dose-response functions $E[Y_i(t)]$ and their derivatives $dE[Y_i(t)]/dt$ are obtained. Standard errors are estimated using 1,000 bootstrap replications \citep{Efron1986, MacKinnon2006}. 




\paragraph{Lottery Winnings and Earnings}
First, the association between the size of lottery winnings and subsequent labor market earnings is re-analyzed using the \cite{Hirano2004} survey data on Megabucks lottery winners in Massachusetts from the mid-1980s.\footnote{The data were originally analyzed by \cite{Imbens2001}.} The dataset contains information on the \textit{prize amount} measured in \$1,000, \textit{labor earnings} six years after winning the lottery, as well as some covariates (\textit{age}, \textit{winning year}, \textit{working status} when winning the lottery, \textit{years of high school}, \textit{years of college}, an indicator for being \textit{male}, the \textit{number of tickets bought} and \textit{previous earnings} in the years one to six prior to winning). While the \textit{prize amount} is randomly assigned, survey and item non-response lead to non-zero correlations of covariates with the treatment variable. The estimation sample consists of $N=201$ lottery winners.\footnote{Compared to \cite{Hirano2004}, a complete case analysis is performed, i.e. individuals with missing data on subsequent labor market earnings were dropped. Moreover, one individual with much lower lottery winnings than the rest of the sample was excluded.} To make the normality assumption used by IPW and CBGPS more credible, the treatment variable T is \textit{log(prize amount)}. The same specification as used in \cite{Hirano2004} is employed to estimate balancing weights.\footnote{Computation times to obtain weights are far below one second for IPW, CBPS, and EBCT. The npCBGPS (GBM) algorithm takes about 3.5 (6.5) seconds to converge. All computations were performed on computer with an 2,7 GHz Intel Core i5 CPU and 8 GB 1600 MHz DDR3 RAM.} An overview of Pearson correlations as well as mean absolute correlations before and after weighting can be found in Table  \ref{bal_lottery}.

\bc
[Insert Table \ref{bal_lottery} about here]
\ec

Raw Pearson correlations range in absolute value from close to zero for the \textit{number of tickets bought} to almost 30\% in the case of the \textit{male} indicator. Correlation coefficients based on the re-weighted samples show that all balancing approaches lead to substantial improvements in overall covariate balance as indicated by the mean absolute correlation.  The non-parametric  CBGPS and EBCT perform best, delivering essentially perfect finite sample balance. GBM yields the smallest overall decline in absolute correlations. For the covariates \textit{years in high school} and \textit{winning year}, it leads to an increase in imbalance after weighting. Table \ref{bal_lottery} also provides the maximum weight assigned to a single individual, which ranges from 1.35\% (EBCT) to 2.94\% (GBM).

\begin{figure}[ht!] \vspace{-3em}
\caption{Empirical Application -- Effects of Lottery Winnings}
\label{est_lottery}
\bc
\includegraphics[scale=0.65, clip, trim=0cm 0cm 0cm 1.8cm]{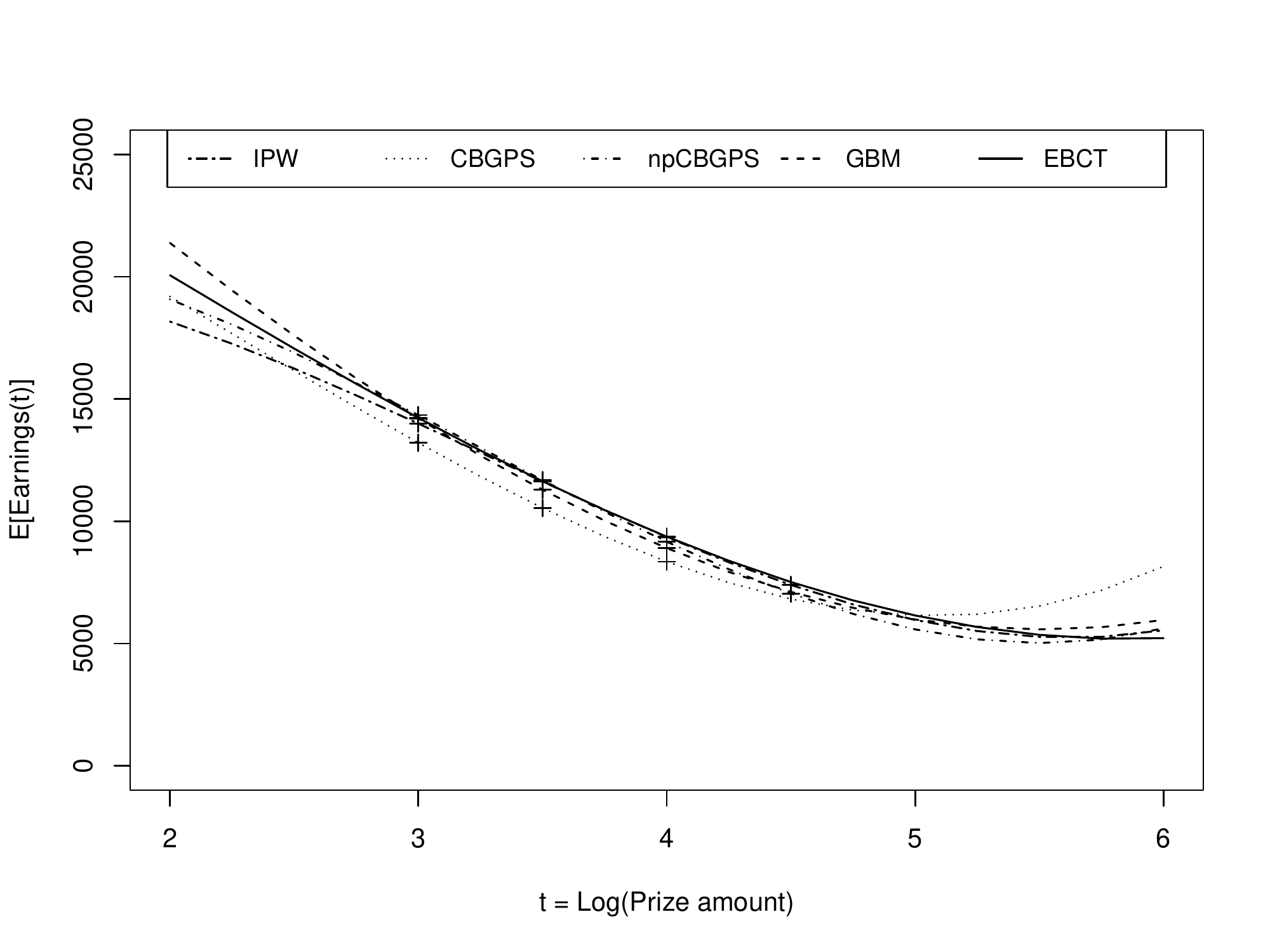}\\
\begin{scriptsize}
\begin{spacing}{1.0}
\parbox{42em}{Note: This graph shows the estimated dose-response function (DRF) between the \textit{log(prize amount)} and subsequent \textit{labor earnings} based on a weighted least squares regression using a cubic specification. Re-weighting approaches employed are  inverse probability weighting estimated via OLS \citep[IPW, see][]{Robins2000}, (non-) parametric covariate balancing generalized propensity scores \citep[np-/CBGPS, see][]{fong2018}, generalized boosted modeling \citep[GBM, see][]{Zhu2015}  as well as the novel entropy balancing for continuous treatments (EBCT).  Values of the DRF are marked with a $+$ if its derivative is significantly different from zero at least at the 10\% level based on bootstrapped standard errors obtained using $R=1,000$ replications.}   \vspace{-7em}
\end{spacing}
\end{scriptsize}
\ec
\end{figure}

\bc
[Insert Figure \ref{est_lottery} about here]
\ec

Figure \ref{est_lottery} shows the estimated DRFs based on the different weighting approaches. In accordance to the results of \cite{Hirano2004}, there is a clear downward-sloping relationship between the \textit{prize amount} and subsequent \textit{labor earnings}. The general patterns obtained across methods are quite similar with slightly more variation in DRF estimates in the tails of the prize distribution. However, derivatives of the DRFs are only significantly different from zero at the 10\% significance level (indicated by a $+$) for \textit{log(prize amount)} in the range of 3 to 4.5 log-points. Hence, despite the differences, none of the slopes of the DRFs are statistically different from zero in the tails.

\paragraph{Smoking Intensity and Medical Expenditures}
Next, the relationship between smoking intensity and medical expenditures is re-visited using data of \cite{Imai2004}, originally analyzed by \cite{JOHNSON2003}. The data stem from the National Medical Expenditure Survey 1987, covering current or previous cigarette smokers and their information on their smoking behavior, (validated) medical expenditures and some background characteristics. Based on the data on past cigarette consumption, the number of \textit{pack years} smoked ($=$ \textit{smoking duration} in years $\cdot$ cigarette \textit{packs smoked per day}) is generated as a measure of smoking intensity. Available background characteristics are the continuous (\textit{starting}) \textit{age}, a \textit{male} indicator and categorical variables on \textit{race}, \textit{seatbelt usage}, \textit{education}, \textit{marital status}, \textit{census region} of residence and \textit{poverty status}. For the estimation of balancing weights, squared and cubed terms in (\textit{starting}) \textit{age} are also included in the specification. The treatment variable is again taken in logarithmic terms to reduce the skewness of the distribution and make the normality assumption by IPW and CBGPS more plausible.\footnote{Computation times to obtain weights are, in ascending order: 0.8 seconds (EBCT), 3.5 seconds (IPW), 8.5 seconds (CBPS), 4.5 minutes (GBM) and 15 minutes (npCBGPS).} The estimation sample consists of $N=9,408$ individuals.\footnote{Compared to the analysis by \cite{Imai2004}, individuals below the 1st percentile and individuals above the 99th percentile of the pack years distribution are dropped as the density of smokers in this region of the distribution is extremely small. Moreover, individuals above the 99th percentile of the medical expenditure distribution are dropped to reduce problems with extreme outliers.} An overview of (mean absolute) Pearson correlations before and after weighting is given by Table \ref{bal_smoking}.

\bc
[Insert Table \ref{bal_smoking} about here]
\ec

Before weighting there is a substantial absolute correlation between \textit{log(pack years)} and the polynomials in \textit{age} (38 to 47\%). Correlations with the other covariates are smaller in magnitude but often substantive nonetheless. Compared to the unweighted sample, IPW increases the mean absolute correlation between covariates and the smoking intensity from 12\% to 16\%, leading to higher correlations in magnitude for starting \textit{age}, the \textit{male} indicator, \textit{seatbelt usage}, \textit{region of residence} and \textit{income}. Clearly, this is a very unfavorable balancing outcome. The parametric CBGPS reduces correlations for almost all variables or leads to slight increases in correlations for variables that were initially almost completely uncorrelated with the treatment. Its non-parametric counterpart is about as successful in reducing absolute correlations for covariates that were initially heavily related to the treatment. However, it leads to larger increases in imbalance for variables with lower initial correlations. For example, the correlation between the \textit{male} indicator and the treatment increases from 14\% to 20\% after weighting using the npCBGPS.  GBM is almost as effective in alleviating correlations between covariates and the treatment as the parametric CBGPS. Only the absolute correlation between the \textit{male} indicator and the treatment of 14\% remains above the 0.1 threshold suggested by \cite{Zhu2015}. In contrast to the other weighting approaches, EBCT yields perfect finite sample balance in terms of correlations also in this application. Moreover, it also produces the smallest maximum weight shares with 0.41\% compared to 0.65\% (GBM), 6.6\% (npCBGPS), IPW (7.1\%) and almost 13\% (CBGPS). Hence, especially the parametric CBGPS method achieves better balance only by allowing for relatively extreme weights in this setting. 

\begin{figure}[ht!] 
\caption{Empirical Application -- Effects of Smoking}
\label{est_smoking}
\bc
\includegraphics[scale=0.65, clip, trim=0cm 0cm 0cm 1.8cm]{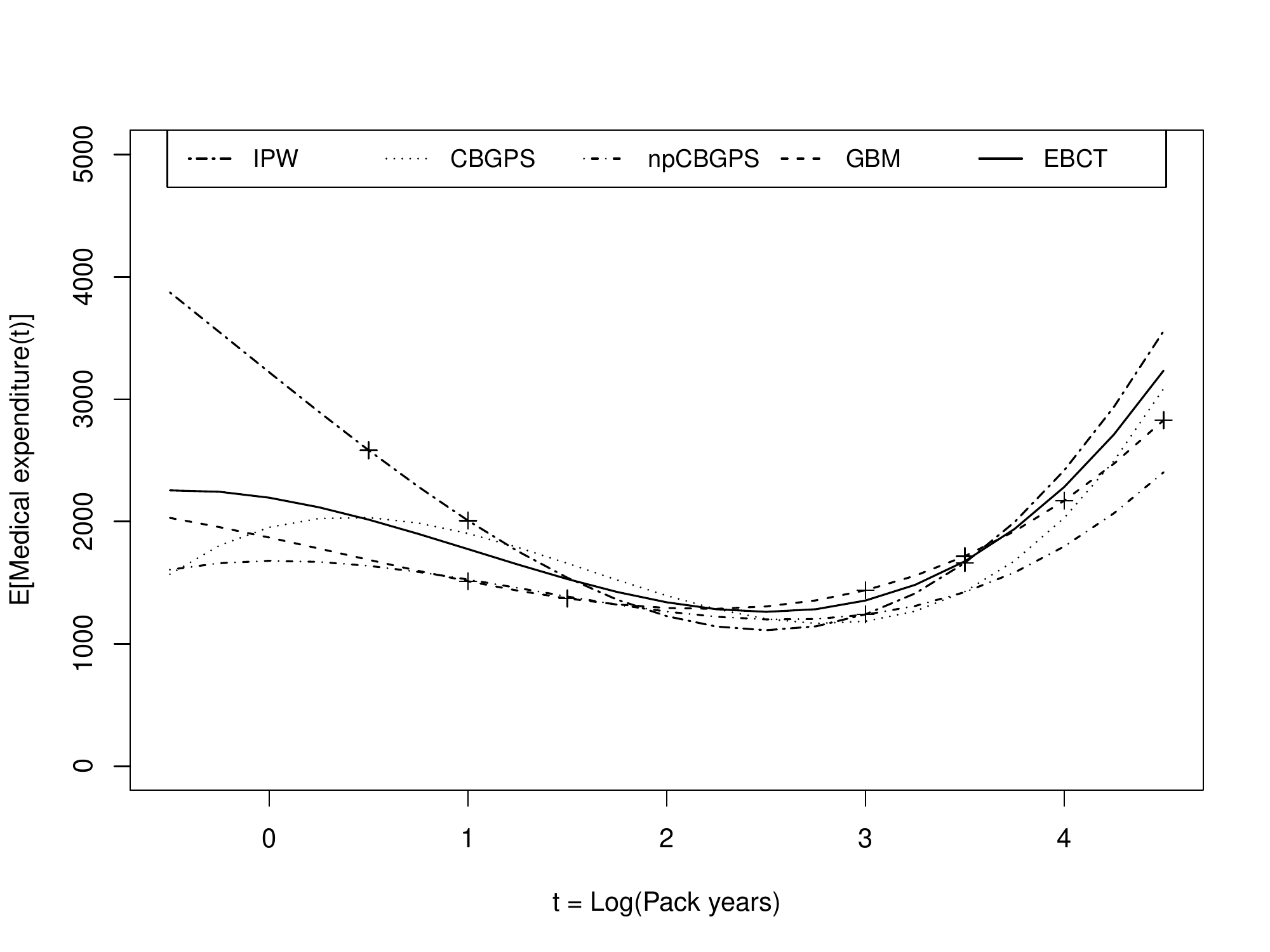}\\
\begin{scriptsize}
\begin{spacing}{1.0}
\parbox{42em}{Note: This graph shows the estimated dose-response function(DRF) between the \textit{log(pack years)} and \textit{medical expenditures} based on a weighted least squares regression using a cubic specification. Re-weighting approaches employed are  inverse probability weighting estimated via OLS \citep[IPW, see][]{Robins2000}, (non-) parametric covariate balancing generalized propensity scores \citep[np-/CBGPS, see][]{fong2018}, generalized boosted modeling \citep[GBM, see][]{Zhu2015}  as well as the novel entropy balancing for continuous treatments (EBCT).  Values of the DRF are marked with a $+$ if its derivative is significantly different from zero at least at the 10\% level based on bootstrapped standard errors obtained using $R=1,000$ replications.} 
\end{spacing}
\end{scriptsize}
\ec
\end{figure}

\bc
[Insert Figure \ref{est_smoking} about here]
\ec

Estimated DRFs are plotted in Figure  \ref{est_smoking}. The estimates suggest a slightly negative relationship between smoking intensity medical expenditures up to about 2.5 \textit{log(pack years)}, after which there is a clear increase in medical expenditures with increased smoking intensity. Again, estimated DRFs differ mostly in the tails of the treatment variable distribution. However, differences are much larger in this application, especially for low levels of the smoking intensity where some estimates suggest significant non-zero derivatives of the DRF.

\section{Conclusions}

This paper extends the entropy balancing methodology for the estimation of covariate balancing weights to the context of  continuous treatments. Owing to its flexibility and globally convex optimization problem, EBCT achieves superior finite sample balance compared to other re-weighting methods based on the GPS. In fact, EBCT eradicates correlations between covariates and the continuous treatment variable even when selection into treatment is strong. At the same time, EBCT effectively avoids assigning extreme weights to single units. Extensive Monte-Carlo simulations show that effect estimates using EBCT display similar or lower bias and uniformly lower RMSE compared to the other weighting methods considered. All in all, these properties make the proposed EBCT method an attractive approach for the estimation of dose-response functions. 


\newpage
\setlength{\baselineskip}{6pt}
\bibliography{litbank_empwifo_v3}
\newpage
\section*{Tables and Figures}

\renewcommand{\baselinestretch}{1.0}
\renewcommand{\arraystretch}{1.2}

\vspace{-1em}
\begin{table}[ht!]
\setlength{\tabcolsep}{3pt}
\begin{center}
\begin{threeparttable}
\caption{Simulation Results - Bias and Root Mean Squared Error ($N=200$)}\label{mc_res}
\begin{footnotesize}
\begin{tabular}{lccccccccccccc}
\hline \hline
& \multicolumn{6}{c}{Moderate Selection $\sigma=4$}& & \multicolumn{6}{c}{Strong Selection $\sigma=2$}\\ \cline{2-7}  \cline{9-14}
Degree of non- & \multicolumn{2}{c}{}& & \multicolumn{2}{c}{} & & \multicolumn{2}{c}{}& & \multicolumn{2}{c}{} \\ 
linearity/additivity &  \multicolumn{2}{c}{None}& \multicolumn{2}{c}{Mild}&  \multicolumn{2}{c}{Moderate} & &  \multicolumn{2}{c}{None}& \multicolumn{2}{c}{Mild}&  \multicolumn{2}{c}{Moderate}  \\  
in $Y|X$&  \multicolumn{2}{c}{$\eta=1$}& \multicolumn{2}{c}{$\eta=1.25$}&  \multicolumn{2}{c}{$\eta=1.5$} & &  \multicolumn{2}{c}{$\eta=1$}& \multicolumn{2}{c}{$\eta=1.25$}&  \multicolumn{2}{c}{$\eta=1.5$}  \\  
\hline
&Bias& RMSE&Bias& RMSE &Bias& RMSE&& Bias& RMSE& Bias& RMSE& Bias& RMSE\\
\cline{2-7}  \cline{9-14} 
\quad Unweighted & 24.8 & 26.1 & 41.5 & 42.7 & 69.4 & 71.2 && 49.0 & 50.3 & \ 81.1 & \ 82.3 & 138.9 & 140.8 \\ 
 \\ \multicolumn{8}{l}{Specification 1: Correctly specified $E[T\mid X]$}\\
  \quad IPW & \ 4.1 & 15.8 & \ 6.1 & 20.1 & \ 8.6 & 26.4 && 18.8 & 37.0 & \ 30.1 & \ 47.8 & \ 50.5 & \ 71.2 \\ 
  \quad CBGPS & \ 5.8 & 13.8 & \ 9.4 & 16.6 & 14.2 & 21.7 && 25.4 &  32.8 & \ 42.5 & \ 48.0 & \ 70.8 & \ 77.1 \\ 
  \quad npCBGPS & \ 0.6 & 12.9 & \ 0.9 & 13.0 & \ 0.7 & 13.2 && \ 4.5 & 31.8 & \ \ 6.7 & \ 33.0 & \ 11.5 & \ 35.9 \\ 
  \quad GBM & \ 9.5 & 15.6 & 14.9 & 21.6 & 24.2 & 31.8 && 30.9 & 42.1 & \ 55.3 & \ 70.4 & 101.0 & 132.1 \\ 
  \quad EBCT & \ 0.4 & 11.9 & \ 0.6 & 11.9 & \ 0.3 & 12.1 && \ 0.6 & 30.8 & \ \ 0.6 & \ 31.8 & \ \ 1.0 & \ 31.6 \\ 

 \\ \multicolumn{8}{l}{Specification 2: Mildly mis-specified $E[T\mid X]$}\\
  \quad   IPW & \ 7.6 & 16.5 & 10.1 & 20.5 & 14.8 & 29.7 && 28.3 & 41.2 & \ 40.4 & \ 53.2 & \ 61.4 & \ 78.9 \\ 
  \quad  CBGPS & \ 8.8 & 15.1 & 12.9 & 18.3 & 19.5 & 25.3 && 33.4 & 39.1 & \ 48.2 & \ 53.2 & \ 76.9 & \ 81.9 \\ 
  \quad  npCBGPS & \ 4.6 & 13.6 & \ 5.8 & 13.9 & \ 7.3 & 15.0 && 19.2 & 36.3 & \ 21.2 & \ 36.6 & \ 27.8 & \ 43.2 \\ 
  \quad  GBM & 10.6 & 16.6 & 16.4 & 22.0 & 26.4 & 34.0 && 35.3 & 43.6 & \ 54.9 & \ 67.6 & \ 96.8 & 119.3 \\ 
  \quad   EBCT & \ 4.7 & 12.9 & \ 5.7 & 13.1 & \ 6.9 & 14.3 && 18.2 & 34.0 & \ 18.8 & \ 33.9 & \ 22.1 & \ 37.3 \\ 
  
 \\ \multicolumn{8}{l}{Specification 3: Strongly mis-specified $E[T\mid X]$}\\
  \quad    IPW & 19.9 & 24.7 & 33.6 & 39.3 & 60.8 & 71.1 && 62.2 & 71.4 & 113.5 & 128.6 & 220.9 & 259.8 \\ 
  \quad    CBGPS & 19.0 & 22.5 & 32.2 & 35.9 & 57.5 & 62.3 && 54.3 & 58.5 & \ 94.6 & \ 99.4 & 173.7 & 183.6 \\ 
  \quad    npCBGPS & 17.9 & 21.9 & 30.8 & 34.5 & 56.0 & 61.2 && 53.9 & 60.3 & \ 94.3 & 100.8 & 169.0 & 179.1 \\ 
  \quad    GBM & 19.9 & 23.7 & 34.0 & 37.9 & 60.2 & 66.8 && 57.6 & 63.2 & \ 99.7 & 107.4 & 189.9 & 208.1 \\ 
  \quad    EBCT & 17.8 & 21.3 & 30.7 & 34.0 & 55.8 & 60.5 & &54.0 & 59.5 & \ 93.4 & \ 99.1 & 168.3 & 177.4 \\

 \hline \hline
\end{tabular}
\end{footnotesize}
\begin{tablenotes}
\begin{scriptsize}
\item Note: This table shows absolute bias and root mean squared error (RMSE) measured in percent of the true treatment effect from the 18 Monte-Carlo simulation scenarios for $N=200$. For each scenario, $R=1,000$ independent replications are performed. Effect estimates are obtained through weighted least squares regression of the outcome $Y$ on the treatment intensity $T$. Re-weighting approaches employed are  inverse probability weighting estimated via OLS \citep[IPW, see][]{Robins2000}, (non-) parametric covariate balancing generalized propensity scores \citep[np-/CBGPS, see][]{fong2018}, generalized boosted modeling \citep[GBM, see][]{Zhu2015}  as well as the novel entropy balancing for continuous treatments (EBCT). 
\item
\end{scriptsize}
\end{tablenotes}
\end{threeparttable}
\end{center}
\end{table}

\begin{table}[ht!]
\begin{center}
\begin{threeparttable}
\caption{Balancing Quality -- Lottery Winnings}\label{bal_lottery}
\begin{footnotesize}
\begin{tabular}{lccccccccccc}
\hline \hline
& \multicolumn{6}{c}{(Weighted) Corr(\textit{log(prize amount)}, $X_k)$}\\ \cline{2-7}
Covariate & Unweighted & IPW & CBGPS & npCBGPS & GBM & EBCT \\ 
  \hline
Age & \ 0.21 & \ 0.06 & 0.02 & 0.00 & \ 0.11 & 0.00 \\ 
  Years in high school & -0.03 &\ 0.00 & 0.00 & 0.00 & -0.07 & 0.00 \\ 
  Years in college & \ 0.04 & \ 0.05 & 0.03 & 0.00 & \ 0.01 & 0.00 \\ 
  Male & \ 0.29 & \ 0.07 & 0.03 & 0.00 & \ 0.12 & 0.00 \\ 
  Tickets bought & \ 0.03 & \ 0.01 & 0.00 & 0.00 & \ 0.02 & 0.00 \\ 
  Working then & \ 0.07 & \ 0.02 & 0.02 & 0.00 & \ 0.03 & 0.00 \\ 
  Winning year & \ 0.03 & -0.01 & 0.01 & 0.00 & \ 0.10 & 0.00 \\ 
  Earnings year-1 & \ 0.13 & \ 0.01 & 0.02 & 0.00 & -0.02 & 0.00 \\ 
  Earnings year-2 & \ 0.19 & \ 0.02 & 0.02 & 0.00 & \ 0.00 & 0.00 \\ 
  Earnings year-3 & \ 0.21 & \ 0.02 & 0.02 & 0.00 & \ 0.01 & 0.00 \\ 
  Earnings year-4 & \ 0.21 & \ 0.02 & 0.03 & 0.00 & \ 0.03 & 0.00 \\ 
  Earnings year-5 & \ 0.16 & \ 0.02 & 0.02 & 0.00 &\  0.00 & 0.00 \\ 
  Earnings year-6 & \ 0.17 & \ 0.06 & 0.04 & 0.00 & \ 0.02 & 0.00 \\ 
\\ Mean absolute correlation& \ 0.14 & \ 0.03 & 0.02 & 0.00 & \ 0.04 & 0.00\\
Maximum weight in \% & & \ 1.68 & 1.68 & 2.39 & \ 2.94 & 1.35\\
   \hline\hline
\end{tabular}
\end{footnotesize}
\begin{tablenotes}
\begin{scriptsize}
\item Note: The table shows Pearson correlations between the treatment variable $t=$\textit{log(prize amount)} and covariates in the raw sample of \cite{Hirano2004} as well as in the re-weighed samples. Re-weighting approaches employed are inverse probability weighting estimated via OLS \citep[IPW, see][]{Robins2000}, (non-) parametric covariate balancing generalized propensity scores \citep[np-/CBGPS, see][]{fong2018}, generalized boosted modeling \citep[GBM, see][]{Zhu2015}  as well as the novel entropy balancing for continuous treatments (EBCT). 

\item
\end{scriptsize}
\end{tablenotes}
\end{threeparttable}
\end{center}
\end{table}

\begin{table}[ht!]
\begin{center}
\begin{threeparttable}
\caption{Balancing Quality -- Smoking Intensity}\label{bal_smoking}
\begin{footnotesize}
\begin{tabular}{lccccccccccc}
\hline \hline
& \multicolumn{6}{c}{(Weighted) Corr(\textit{log(pack years)}, $X_k)$}\\ \cline{2-7}
Covariate $X_k$ & Unweighted & IPW & CBGPS & npCBGPS & GBM & EBCT \\ 
  \hline
Starting age\\ \quad Linear & -0.14 & \ 0.20 & -0.07 & -0.06 & -0.08 & \ 0.00 \\ 
  \quad Squared &-0.13 & \ 0.19 & -0.04 & -0.05 & -0.07 & \ 0.00 \\
  \quad Cubed & -0.11 & \ 0.16 & -0.02 & -0.04 & -0.06 & \ 0.00 \\ 

  Age\\ \quad Linear  &\ 0.47 & -0.34 & \ 0.06 & \ 0.07 & \ 0.01 & \ 0.00 \\ 
  \quad Squared & \ 0.43 & -0.32 & \ 0.06 & \ 0.07 & \ 0.01 & \ 0.00 \\ 
  \quad Cubed  & \ 0.38 & -0.29 & \ 0.05 & \ 0.08 & \ 0.01 & \ 0.00 \\ 
 
  Male & \ 0.14 & \ 0.23 &\  0.04 & \ 0.20 & \ 0.14 & \ 0.00 \\  

 Race\\ \quad Black & -0.14 & \ 0.19 & -0.01 & -0.06 & -0.01 & \ 0.00 \\ 
 \quad  Other & \ 0.19 & -0.24 & \ 0.02 & -0.10 & \ 0.03 & \ 0.00 \\ 

  Seatbelt usage\\ \quad  Sometimes &  -0.01 & \ 0.2 & -0.03 & -0.03 & \ 0.05 & \ 0.00 \\ 
  \quad Often & -0.04 & -0.18 & \ 0.05 & \ 0.02 & -0.11 & \ 0.00 \\ 

  Education\\ \quad High school & \ 0.01 & -0.04 &\  0.02 & \ 0.10 &\  0.06 & \ 0.00 \\ 
  \quad Some college & -0.04 & -0.02 & -0.03 & -0.01 & \ 0.04 & \ 0.00 \\ 
  \quad College degree & \ 0.11 & -0.03 & \ 0.00 & -0.10 & -0.01 & \ 0.00 \\ 
  \quad Other & \ 0.11 & -0.18 & -0.01 &\  0.02 & -0.02 & \ 0.00 \\ 

Marital status\\ \quad  Widowed &  \ 0.04 & \ 0.07 & -0.02 & \ 0.13 & \ 0.05 & \ 0.00 \\ 
\quad  Separated & -0.02 & \ 0.11 & -0.02 & -0.03 & \ 0.04 & \ 0.00 \\ 
\quad  Never married & -0.26 & \ 0.17 & \ 0.07 & -0.09 & -0.01 & \ 0.00 \\ 
Census region\\ \quad  Mid-west & \ 0.02 & \ 0.16 & -0.03 & \ 0.07 & \ 0.05 & \ 0.00 \\ 
 \quad  South & -0.02 & -0.20 & \ 0.05 & -0.07 & -0.08 & \ 0.00 \\ 
 \quad  West & -0.01 & \ 0.15 & -0.02 & -0.02 & \ 0.01 & \ 0.00 \\ 

 Poverty status\\ \quad Poor & -0.02 & -0.03 & -0.02 & \ 0.10 & -0.01 &\  0.00 \\ 
\quad  Low income &-0.01 & -0.04 & -0.01 & -0.09 & \ 0.00 & \ 0.00 \\ 
 \quad  Middle income & \ 0.00 & -0.15 & -0.03 & \ 0.10 & -0.01 & \ 0.00 \\ 
 \quad  High income & \ 0.03 & \ 0.18 & -0.02 & -0.02 & \ 0.00 & \ 0.00 \\  
\\ Mean absolute correlation& \ 0.12 & \ 0.16 & \ 0.03 &\  0.08 & \ 0.04 & \ 0.00\\
Maximum weight in \% &  & \ 7.1 & \ 12.9 & \ 6.6 & \ 0.7 & \ 0.4 \\ 

   \hline\hline
\end{tabular}
\end{footnotesize}
\begin{tablenotes}
\begin{scriptsize}
\item Note: The table shows absolute Pearson correlations between the treatment variable $t=$\textit{ln(pack years)} and covariates in the raw sample of \cite{Imai2004} as well as in the re-weighed samples. Re-weighting approaches employed are  inverse probability weighting estimated via OLS \citep[IPW, see][]{Robins2000}, (non-) parametric covariate balancing generalized propensity scores \citep[np-/CBGPS, see][]{fong2018}, generalized boosted modeling \citep[GBM, see][]{Zhu2015}  as well as the novel entropy balancing for continuous treatments (EBCT). 

\item
\end{scriptsize}
\end{tablenotes}
\end{threeparttable}
\end{center}
\end{table}

\clearpage

\clearpage

\clearpage


\renewcommand{\thesection}{A}
\renewcommand{\thetable}{A.\arabic{table}}
\setcounter{table}{0}
\renewcommand{\thefigure}{A.\arabic{figure}}
\setcounter{figure}{0}

\renewcommand{\baselinestretch}{1.0}
\renewcommand{\arraystretch}{1.0}

\section*{Appendix A}

\begin{table}[ht!]
\setlength{\tabcolsep}{3pt}
\begin{center}
\begin{threeparttable}
\caption{Simulation Results - Bias and Root Mean Squared Error ($N=500$)}\label{mc_res500}
\begin{footnotesize}
\begin{tabular}{lccccccccccccc}
\hline \hline
& \multicolumn{6}{c}{Moderate Selection $\sigma=4$}& & \multicolumn{6}{c}{Strong Selection $\sigma=2$}\\ \cline{2-7}  \cline{9-14}
Degree of non- & \multicolumn{2}{c}{}& & \multicolumn{2}{c}{} & & \multicolumn{2}{c}{}& & \multicolumn{2}{c}{} \\ 
linearity/additivity &  \multicolumn{2}{c}{None}& \multicolumn{2}{c}{Mild}&  \multicolumn{2}{c}{Moderate} & &  \multicolumn{2}{c}{None}& \multicolumn{2}{c}{Mild}&  \multicolumn{2}{c}{Moderate}  \\  
in $Y|X$&  \multicolumn{2}{c}{$\eta=1$}& \multicolumn{2}{c}{$\eta=1.25$}&  \multicolumn{2}{c}{$\eta=1.5$} & &  \multicolumn{2}{c}{$\eta=1$}& \multicolumn{2}{c}{$\eta=1.25$}&  \multicolumn{2}{c}{$\eta=1.5$}  \\  
\hline
&Bias& RMSE&Bias& RMSE &Bias& RMSE&& Bias& RMSE& Bias& RMSE& Bias& RMSE\\
\cline{2-7}  \cline{9-14} 
  \quad Unweighted & \ 24.6 & \ 25.2 & \ 41.4 & \ 41.9 & \ 69.4 & \ 70.2 && \ 49.3 & \ 49.8 & \ 81.7 & \ 82.2 & 139.6 & 140.4 \\ 
 \\ \multicolumn{8}{l}{Specification 1: Correctly specified $E[T\mid X]$}\\
  \quad    IPW &\ \  1.7 & \ 12.3 & \ \ 4.5 & \ 15.2 & \ \ 5.4 & \ 22.3 && \ 16.5 & \ 32.3 & \ 26.6 & \ 42.0 & \ 46.4 & \ 64.6 \\ 
   \quad   CBGPS & \ \ 2.5 & \ \ 9.1 & \ \ 5.1 & \ 10.9 & \ \ 7.4 & \ 14.1 && \ 22.5 & \ 28.2 & \ 36.3 & \ 40.5 & \ 61.4 &\  66.3 \\ 
   \quad   npCBGPS & \ \ 0.1 & \ \ 9.0 & \ \ 0.5 & \ \ 9.3 & \ \ 0.4 & \ \ 9.3 && \ \ 2.5 & \ 25.7 & \ \ 3.3 & \ 26.1 & \ \ 6.0 &\  27.2 \\ 
   \quad   GBM & \ \ 4.7 & \ 10.3 & \ \ 9.0 & \ 14.0 & \ 13.5 & \ 19.5 && \ 21.5 & \ 30.9 & \ 36.8 & \ 49.3 & \ 64.2 & \ 79.1 \\ 
    \quad  EBCT & \ \ 0.0 & \ \ 7.4 &\ \  0.6 & \ \ 7.7 & \ \ 0.1 & \ \ 7.5 && \ \ 0.6 & \ 19.3 & \ \ 0.1 & \ 18.9 & \ \ 0.4 & \ 20.2 \\ 
 
 \\ \multicolumn{8}{l}{Specification 2: Mildly mis-specified $E[T\mid X]$}\\
    \quad  IPW & \ \ 6.6 & \ 12.7 & \ \ 8.3 & \ 16.6 & \ 11.0 & \ 22.0 && \ 25.7 & \ 36.1 & \ 35.8 & \ 48.0 & \ 55.7 & \ 68.4 \\ 
    \quad  CBGPS & \ \ 6.7 & \ 10.8 & \ \ 8.6 & \ 12.7 & \ 12.5 & \ 16.8 && \ 30.1 & \ 34.1 & \ 43.8 & \ 47.8 & \ 69.3 & \ 73.1 \\ 
    \quad  npCBGPS & \ \ 4.9 & \ 10.0 & \ \ 5.8 & \ 10.7 & \ \ 7.0 & \ 11.4 && \ 16.7 & \ 28.4 & \ 19.5 & \ 30.1 & \ 24.5 & \ 33.4 \\ 
    \quad  GBM & \ \ 8.2 & \ 11.9 & \ 11.5 & \ 15.3 & \ 16.7 & \ 21.1 && \ 28.5 & \ 35.3 & \ 42.1 & \ 50.5 & \ 66.5 & \ 79.3 \\ 
    \quad  EBCT & \ \ 5.2 & \ \ 9.0 & \ \ 5.6 & \ \ 9.4 & \ \ 6.6 & \ \ 9.9 && \ 17.2 & \ 25.0 & \ 19.7 & \ 26.7 & \ 22.7 & \ 28.9 \\ 

 \\ \multicolumn{8}{l}{Specification 3: Strongly mis-specified $E[T\mid X]$}\\
    \quad  IPW & \ 19.7 & \ 22.5 & \ 34.5 & \ 37.4 & \ 62.0 & \ 72.5 && \ 71.6 & \ 80.2 & 128.8 & \ 142.8 & 263.8 & 306.0 \\ 
    \quad  CBGPS & \ 18.7 & \ 20.5 & \ 32.7 & \ 34.6 & \ 57.4 & \ 60.7 && \ 59.2 & \ 62.0 & 104.9 & 108.5 & 198.7 & 207.7 \\ 
    \quad  npCBGPS & \ 18.5 & \ 20.5 & \ 32.1 & \ 34.1 & \ 56.7 & \ 60.3 && \ 56.6 & \ 59.9 & 100.2 & 103.8 & 186.5 & 193.8 \\ 
    \quad  GBM & \ 19.5 & \ 21.7 & \ 33.7 & \ 35.9 & \ 59.6 & \ 64.1 && \ 63.1 & \ 67.1 & 113.9 & 121.2 & 217.3 & 236.7 \\ 
    \quad  EBCT & \ 18.3 & \ 19.8 & \ 31.7 & \ 33.2 & \ 55.9 & \ 58.3 && \ 56.1 & \ 58.3 & \ 97.3 & \ 99.8 & 179.5 & 184.7 \\ 
 \hline \hline
\end{tabular}
\end{footnotesize}
\begin{tablenotes}
\begin{scriptsize}
\item Note: This table shows absolute bias and root mean squared error (RMSE) measured in percent of the true treatment effect from the 18 Monte-Carlo simulation scenarios for $N=500$. For each scenario, $R=1,000$ independent replications are performed. Effect estimates are obtained through weighted least squares regression of the outcome $Y$ on the treatment intensity $T$. Re-weighting approaches employed are  inverse probability weighting estimated via OLS \citep[IPW, see][]{Robins2000}, (non-) parametric covariate balancing generalized propensity scores \citep[np-/CBGPS, see][]{fong2018}, generalized boosted modeling \citep[GBM, see][]{Zhu2015}  as well as the novel entropy balancing for continuous treatments (EBCT). 
\item
\end{scriptsize}
\end{tablenotes}
\end{threeparttable}
\end{center}
\end{table}

\begin{table}[ht!]
\setlength{\tabcolsep}{3pt}
\begin{center}
\begin{threeparttable}
\caption{Simulation Results - Bias and Root Mean Squared Error ($N=1,000$)}\label{mc_res1000}
\begin{footnotesize}
\begin{tabular}{lccccccccccccc}
\hline \hline
& \multicolumn{6}{c}{Moderate Selection $\sigma=4$}& & \multicolumn{6}{c}{Strong Selection $\sigma=2$}\\ \cline{2-7}  \cline{9-14}
Degree of non- & \multicolumn{2}{c}{}& & \multicolumn{2}{c}{} & & \multicolumn{2}{c}{}& & \multicolumn{2}{c}{} \\ 
linearity/additivity &  \multicolumn{2}{c}{None}& \multicolumn{2}{c}{Mild}&  \multicolumn{2}{c}{Moderate} & &  \multicolumn{2}{c}{None}& \multicolumn{2}{c}{Mild}&  \multicolumn{2}{c}{Moderate}  \\  
in $Y|X$&  \multicolumn{2}{c}{$\eta=1$}& \multicolumn{2}{c}{$\eta=1.25$}&  \multicolumn{2}{c}{$\eta=1.5$} & &  \multicolumn{2}{c}{$\eta=1$}& \multicolumn{2}{c}{$\eta=1.25$}&  \multicolumn{2}{c}{$\eta=1.5$}  \\  
\hline
&Bias& RMSE&Bias& RMSE &Bias& RMSE&& Bias& RMSE& Bias& RMSE& Bias& RMSE\\
\cline{2-7}  \cline{9-14} 
  \quad Unweighted & \ 24.7 & \ 25.0 & \ 41.0 & \ 41.2 & \ 69.6 & \ 69.9 &&  \ 49.2 & \ 49.4 & \ 81.8 & \ 82.1 & 139.5 & 139.9 \\ 

 \\ \multicolumn{8}{l}{Specification 1: Correctly specified $E[T\mid X]$}\\
  \quad   IPW & \ \ 1.1 & \ 10.6 & \ \ 2.4 & \  12.1 & \ \ 4.9 & \ 16.8 && \ 13.7 & \ 26.8 & \ 22.4 & \ 38.5 & \ 36.7 & \ 58.3 \\ 
  \quad   CBGPS & \ \  1.1 & \ \ 6.9 & \ \ 1.7 & \ \ 7.3 & \ \ 3.7 & \ \ 9.5 && \ 18.6 & \ 23.7 & \ 31.6 & \ 36.0 & \ 52.8 & \ 57.3 \\ 
  \quad   npCBGPS & \ \ 0.1 & \ \ 7.5 & \ \ 0.1 & \ \ 7.4 & \ \ 0.8 & \ \ 7.8 && \ \ 1.0 & \ 22.8 & \ \ 2.7 & \  22.8 & \ \ 4.0 & \ 26.0 \\ 
  \quad   GBM & \ \ 3.6 & \ \ 8.0 & \ \ 5.8 & \ \ 9.7 & \ 10.2 & \ 14.8 && \ 17.2 & \ 25.2 & \ 29.3 & \ 37.3 & \ 50.9 & \ 75.1 \\ 
  \quad   EBCT & \ \ 0.1 & \ \ 5.4 & \ \ 0.0 & \ \ 5.3 & \ \ 0.4 & \ \ 5.5 && \ \ 0.8 & \ 14.3 & \ \ 0.2 & \ 14.3 & \ \ 0.6 & \ 14.4 \\ 
  
 \\ \multicolumn{8}{l}{Specification 2: Mildly mis-specified $E[T\mid X]$}\\
  \quad   IPW & \ \ 6.0 & \ 10.1 & \ \ 7.2 & \ 13.6 & \ \ 9.8 & \ 23.3 && \ 24.8 & \ 33.5 & \ 33.8 & \ 42.5 & \ 49.3 & \ 65.0 \\ 
  \quad   CBGPS & \ \ 5.7 & \ \ 8.4 & \ \ 6.8 & \ \ 9.5 & \ \ 8.8 & \ 11.9 && \ 27.8 & \ 31.3 & \ 39.9 & \ 42.9 & \ 61.5 & \ 65.0 \\ 
  \quad   npCBGPS & \ \ 5.2 & \ \ 8.9 & \ \ 5.6 & \ \ 9.1 & \ \ 6.7 & \ 10.1 && \ 16.7 & \ 26.8 & \ 19.4 & \ 27.7 & \ 23.2 & \ 31.6 \\ 
  \quad   GBM & \ \ 7.1 & \ \ 9.9 & \ \ 9.4 & \  11.8 &  \ 12.8 & \ 16.2 && \ 26.3 & \ 31.7 & \ 35.3 & \ 40.1 & \ 53.1 & \ 62.8 \\ 
  \quad   EBCT & \ \ 5.1 & \ \ 7.3 & \ \ 5.6 & \ \ 7.7 & \ \ 6.4 & \ \ 8.4 && \ 17.5 & \ 21.8 & \ 19.8 & \ 23.5 & \ 22.0 & \ 25.7 \\ 

 \\ \multicolumn{8}{l}{Specification 3: Strongly mis-specified $E[T\mid X]$}\\
  \quad   IPW & \ 19.7 & \ 21.3 & \ 34.5 & \ 36.9 & \ 61.7 & \ 66.4 && \ 73.5 & \ 80.8 & 146.0 & 162.1 & 288.1 & 326.2 \\ 
  \quad   CBGPS & \ 18.8 & \ 19.8 & \ 33.0 & \ 34.1 & \ 58.7 & \ 60.5 && \ 62.6 & \ 65.0 & 115.3 & 118.9 & 215.0 & 222.8 \\ 
  \quad   npCBGPS & \ 18.9 & \ 20.2 & \ 32.7 & \ 34.2 & \ 58.2 & \ 60.4 && \ 58.1 & \ 60.9 & 105.8 & 109.0 & 190.5 & 196.5 \\ 
  \quad   GBM & \ 19.5 & \ 20.5 & \ 33.4 & \ 34.9 & \ 60.2 & \ 62.9 && \ 65.8 & \ 69.1 & 122.4 & 128.2 & 240.0 & 255.4 \\ 
  \quad   EBCT & \ 18.5 & \ 19.2 & \ 31.8 & \ 32.6 & \ 56.4 & \ 57.6 && \ 56.7 & \ 58.1 &  101.2 &  102.8 & 181.0 & 184.5 \\ 
 
\hline \hline
\end{tabular}
\end{footnotesize}
\begin{tablenotes}
\begin{scriptsize}
\item Note: This table shows absolute bias and root mean squared error (RMSE) measured in percent of the true treatment effect from the 18 Monte-Carlo simulation scenarios for $N=1,000$. For each scenario, $R=1,000$ independent replications are performed. Effect estimates are obtained through weighted least squares regression of the outcome $Y$ on the treatment intensity $T$. Re-weighting approaches employed are  inverse probability weighting estimated via OLS \citep[IPW, see][]{Robins2000}, (non-) parametric covariate balancing generalized propensity scores \citep[np-/CBGPS, see][]{fong2018}, generalized boosted modeling \citep[GBM, see][]{Zhu2015}  as well as the novel entropy balancing for continuous treatments (EBCT). 
\item
\end{scriptsize}
\end{tablenotes}
\end{threeparttable}
\end{center}
\end{table}

\newpage
\setlength{\baselineskip}{8pt}

\renewcommand{\thesection}{B}
\renewcommand{\thetable}{B.\arabic{table}}


\setcounter{table}{0}
\renewcommand{\thefigure}{B.\arabic{figure}}
\setcounter{figure}{0}

\clearpage

\linespread{1.5}\selectfont
\section*{Appendix B: Additional Application}\label{B}

This appendix provides an additional empirical application of the EBCT methodology using data kindly provided by \cite{Mitze2015}. They evaluate the effects of the largest German place-based regional development subsidy (RDS) program (``Gemeinschaftsaufgabe Verbesserung der regionalen Wirtschaftsstruktur'') on regional productivity growth. Since German re-unification in 1990, the program granted subsidies in excess of \EUR 60 billion to lagging regions.  

The data stem from the Federal statistics agency and they are measured at the county-level. The dataset contains information on counties' \textit{subsidy receipt per capita}, \textit{regional productivity growth} and several covariates (\textit{lagged productivity}, \textit{lagged labor productivity growth}, \textit{lagged employment}, \textit{lagged employment growth}, \textit{local investment intensity}, \textit{average firm size}, \textit{turnover from exports}, \textit{human capital}, \textit{population density}, a \textit{net-migration indicator}, an \textit{urban indicator}, information on \textit{settlement structure} and \textit{time dummies}). The period of observation covers the years 1993-2008. All variables are measured in 3-year intervals. In total, there are 869 treated county-period observations. The treatment variable has been Box-Cox transformed with $\lambda\approx 0.15$ to reduce its skewness. For the estimation of balancing weights, the same specification as in  \cite{Mitze2015} is used.

Table \ref{bal_grw} displays the correlations between covariates and the treatment variable before and after weighting. Before weighting, there is a substantial negative correlation of about -0.6 between \textit{lagged labor productivity} and treatment. Associations of similar magnitude but of opposite sign are given for \textit{lagged productivity growth} and \textit{human capital} endowment. This suggests that highly subsidized regions would have performed better than other regions even without the subsidy due to catch up growth. Neglecting these differences, or failing to achieve balance, is thus likely to overstate the effects of the subsidy on regional development. To adjust for this divergence in pre-treatment characteristics, balancing weights are estimated.\footnote{Computation times are, in ascending order, 0.05 seconds (IPW), 0.25 seconds (CBGPS), 0.6 seconds (EBCT), 22 seconds (GBM) and 1.6 minutes (npCBGPS).} All approaches reduce mean absolute correlations between covariates and the treatment variable. However, only EBCT can reduce these correlations to zero. For all other approaches, there remain sizable correlations, especially with respect to \textit{lagged labor productivity} (\textit{growth}). This lack of balance was also documented by  \cite{Mitze2015} in their original analysis using the Hirano-Imbens approach. Similar to the other empirical applications presented before, EBCT puts the least maximum weight on a single unit with about 5.4\%.


\begin{table}[ht!]
\begin{center}
\begin{threeparttable}
\caption{Balancing Quality -- Regional development}\label{bal_grw}
\begin{footnotesize}
\begin{tabular}{lccccccccccc}
\hline \hline
& \multicolumn{6}{c}{(Weighted) Corr((\textit{Subsidy amount}$^\lambda-1$)/$\lambda$, $X_k)$}\\ \cline{2-7}
Covariate $X_k$ & Unweighted & IPW & CBGPS & npCBGPS & GBM & EBCT \\ 
  \hline
Log(lagged labor productivity) & -0.62 & -0.26 & -0.27 & -0.06 & -0.51 & \ 0.00 \\ 
  Lagged labor prod. growth & \ 0.52 & \ 0.33 & \ 0.34 & \ 0.31 & \ 0.11 & \ 0.00 \\ 
  Log(lagged employment) & -0.05 & \ 0.00 & \ 0.00 & -0.08 & -0.02 & \ 0.00 \\ 
  Lagged employment growth & -0.06 & -0.02 & -0.06 & -0.16 & \ 0.13 & \ 0.00 \\ 
  Log(investment intensity) & \ 0.50 & \ 0.19 & \ 0.21 & \ 0.06 & \ 0.18 & \ 0.00 \\ 
  Log(average firm size) & -0.37 & -0.19 & -0.11 & -0.02 & -0.14 & \ 0.00 \\ 
  Log(foreign turnover) & -0.40 & -0.14 & -0.11 & -0.04 & -0.22 & \ 0.00 \\ 
  Log(share manufacturing sector) & -0.42 & -0.25 & -0.16 & -0.01 & -0.23 & \ 0.00 \\ 
  Log(human capital) & \ 0.52 & \ 0.27 & \ 0.20 & \ 0.03 & \ 0.25 & \ 0.00 \\ 
  Log(population density) & -0.16 & -0.05 & -0.06 & \ 0.06 & \ 0.09 & \ 0.00 \\ 
  Net-migration indicator & -0.29 & -0.2 & -0.15 & -0.10 & -0.14 & \ 0.00 \\ 
  Urban indicator & \ 0.07 & -0.02 & \ 0.01 & \ 0.08 & -0.15 & \ 0.00 \\ 
  Settlement structure & \ 0.21 & \ 0.01 & \ 0.04 & \ 0.07 & -0.10 & \ 0.00 \\ 
  time dummy 1 & \ 0.01 & \ 0.05 & \ 0.04 & -0.10 & \ 0.10 & \ 0.00 \\ 
  time dummy 2 & \ 0.06 & \ 0.07 & \ 0.05 & -0.03 & \ 0.12 & \ 0.00 \\ 
  time dummy 3 & -0.01 & -0.02 & -0.02 & -0.04 & \ 0.07 & \ 0.00 \\ 
  time dummy 4 & -0.05 & \ 0.02 & -0.01 & \ 0.04 & -0.17 & \ 0.00 \\ 
\\  Mean Absolute Correlation & \ 0.25 & \ 0.12 & \ 0.11 & \ 0.08 & \ 0.16 & \ 0.00 \\ 
  Maximum weight in \% &  & \ 5.92 & \ 17.5 & \ 28.7 & \ 9.35 & \ 5.44 \\ 
   \hline\hline
\end{tabular}
\end{footnotesize}
\begin{tablenotes}
\begin{scriptsize}
\item Note: The table shows absolute Pearson correlations between the treatment variable $t=$(\textit{Subsidy amount}$^\lambda-1$)/$\lambda$ with $\lambda=$ and covariates in the raw sample of \cite{Mitze2015} as well as in the re-weighed samples. Re-weighting approaches employed are  inverse probability weighting estimated via OLS \citep[IPW, see][]{Robins2000}, (non-) parametric covariate balancing generalized propensity scores \citep[np-/CBGPS, see][]{fong2018}, generalized boosted modeling \citep[GBM, see][]{Zhu2015}  as well as the novel entropy balancing for continuous treatments (EBCT). 
\item
\end{scriptsize}
\end{tablenotes}
\end{threeparttable}
\end{center}
\end{table}

\begin{figure}[ht!] 
\caption{Empirical Application -- Regional Development}
\label{est_grw}
\bc
\includegraphics[scale=0.65, clip, trim=0cm 0cm 0cm 1.8cm]{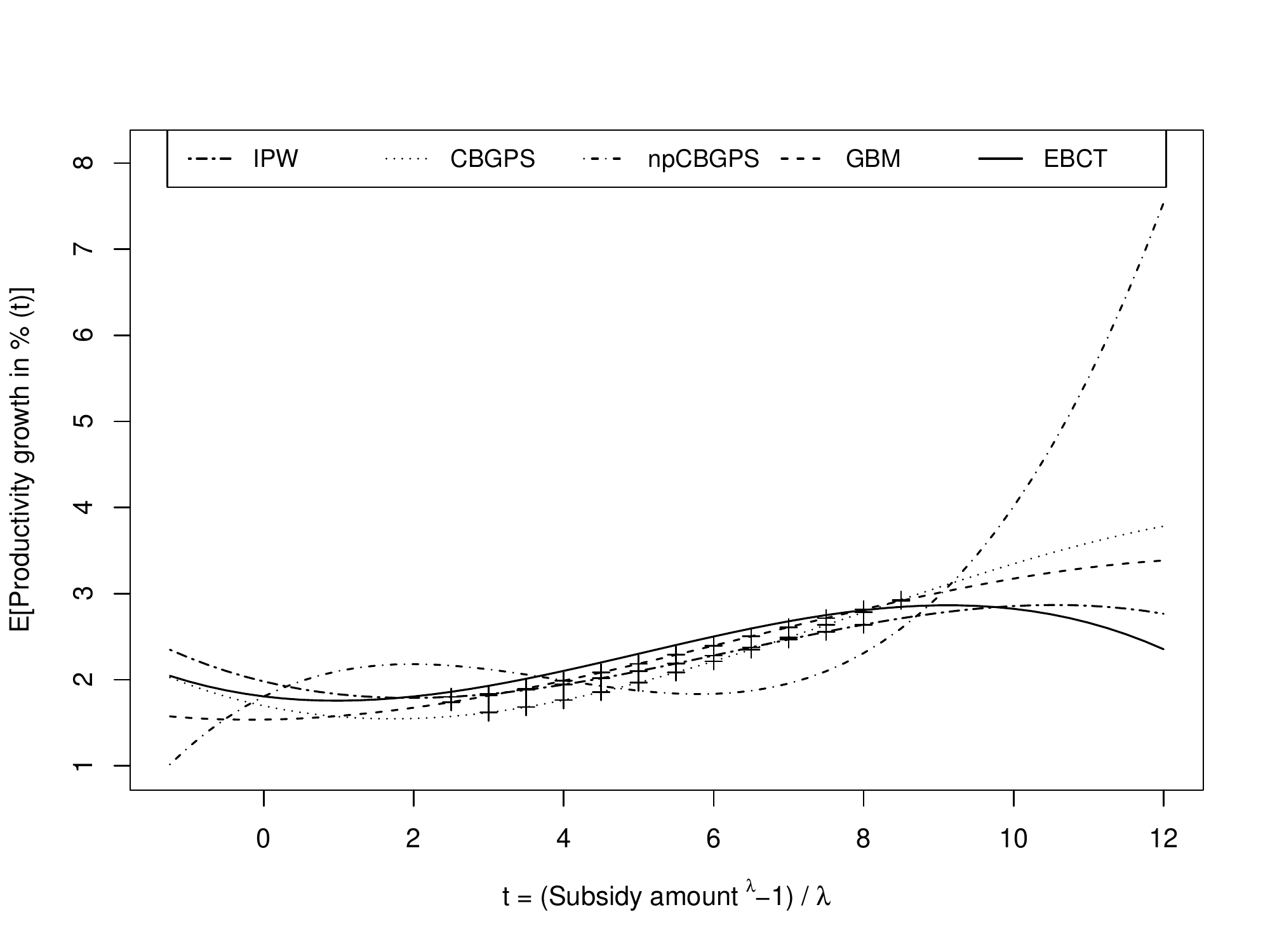}\\
\begin{scriptsize}
\begin{spacing}{1.0}
\parbox{42em}{Note: This graph shows the estimated dose-response function(DRF) between the $t=$(\textit{Subsidy amount}$^\lambda-1$)/$\lambda$ with $\lambda=$ and \textit{productivity growth} based on a weighted least squares regression using a cubic specification. Re-weighting approaches employed are  inverse probability weighting estimated via OLS \citep[IPW, see][]{Robins2000}, (non-) parametric covariate balancing generalized propensity scores \citep[np-/CBGPS, see][]{fong2018}, generalized boosted modeling \citep[GBM, see][]{Zhu2015}  as well as the novel entropy balancing for continuous treatments (EBCT).  Values of the DRF are marked with a $+$ if its derivative is significantly different from zero at least at the 10\% level based on bootstrapped standard errors obtained using $R=1,000$ replications.} 
\end{spacing}
\end{scriptsize}
\ec
\end{figure}

\noindent Estimated DRFs are plotted in Figure  \ref{est_grw}. Most estimates suggest a positive relationship between the subsidy and regional productivity growth for medium values of the treatment intensity. IPW weights yield unrealistic estimates with excessively large gains from the subsidy in the upper tail of the distribution. While point estimates are relatively similar across regressions based on (np-)CBGPS, GBM and EBCT, none of the derivatives of the DRF using EBCT are statistically significant at the 10\% level. This contrasts with the findings of \cite{Mitze2015} who report significantly positive derivatives of the DRF for treatment intensities around the center of the distribution. As EBCT estimates are the most credible due to their superior balancing quality, one should be skeptical of the validity of these findings.

\end{document}